\documentstyle[aps,twocolumn]{revtex}
\begin{document}

\twocolumn[\hsize\textwidth\columnwidth\hsize\csname @twocolumnfalse\endcsname

\title{Some Recent Issues in Quantum Magnetism} 
\author{ R. R. P. Singh, W. E. Pickett, D. W. Hone and D. J. Scalapino}
\address{Institute for Theoretical Physics, University of California,
Santa Barbara, CA 93106}
\maketitle
\widetext

\begin{center}
{\it Summary of ``Quantum Magnetism'' Conference}\\

  {\it Institute for Theoretical Physics, Santa Barbara, California,
        August 16-20, 1999}

\end{center}


\narrowtext
]

\section{Magnetism: Old and New}

Magnetism is an old subject, but it is still full of surprises. It is
deeply rooted in experimental phenomena, yet  it has also been a perennial
source for new theoretical ideas.
It is a remarkably rich area in terms of variety of possible new phases, 
critical phenomena, symmetries of the order parameter,
effective dimensionality and variability of experimental control parameters.
It has been one of the backbones of modern technology, yet
it promises still more technological marvels to come --
ranging from nanostorage devices and spin-electronics to 
quantum computing.
The interplay of magnetism with other solid-state phenomena such as
superconductivity, structural phase transitions and colossal magnetoresistance,
has spurred wide ranging research activities in recent years. 
 
The subject of {\it quantum} magnetism dates
back to the invention of quantum mechanics itself -- starting from
the works of Heisenberg, Bethe and others in the 1920s. 
One of the reasons for the longstanding prominence of
the field in theoretical physics is the existence of simple
models, which are tractable and yet display detailed quantitative
correspondence with real systems.
Recent developments in the synthesis of complex materials and in
the sophistication and quantitative accuracy of experimental probes,
ranging from neutrons to x-rays, optics and
NMR, and the development of new techniques such as the use of
polarized synchrotron radiation,
have reinvigorated the field from an experimental point of view. 
Recent applications of new theoretical techniques, including Bosonization
and conformal field theory, as well as the dramatic improvements in
computational techniques ranging from first principles density functional
calculations to quantum Monte Carlo simulations and density
matrix renormalization group for many-body systems, have opened up
the possibility that properties of complex materials can be theoretically
predicted -- leading to new phenomena and applications.   

It may not be an exaggeration to say that 
{\it the} central problem in condensed matter physics
in recent years has been high temperature superconductivity in
the cuprates, and research in quantum magnetism has had a
big boost from the fact that the stoichiometric parent compounds
of the high temperature superconducting materials are excellent
realizations of quasi-two-dimensional quantum antiferromagnets.
Furthermore, even in the superconducting materials there is
evidence for spin-fluctuations, spin-gaps and spin-stripes,
which are obviously related to quantum magnetism.
However, high temperature superconducting materials now comprise only
a small subfield of the growing activity in quantum magnetism.

Recent advances in the field of magnetism include
experimental realizations of spin chains and ladders, 
inorganic spin-Peierls materials, materials exhibiting colossal 
magnetoresistance (CMR), superconducting ferromagnets,
organic magnetic materials, nanocrystalline magnetic 
materials, molecular magnets,
and artificial structures -- notably, on the mesoscopic scale -- which 
employ magnetic properties to build novel electronic devices. 
These novel phenomena are based on structural complexity that leads
to exchange interactions of varying symmetries and spatial ranges.
In many systems exchange interactions compete, resulting in magnetic
frustration.  Additionally, low dimensional behavior is particularly
accessibility in practice 
in spin systems, because 
exchange interactions usually are short range.  Analogous isolation
in mechanical or electrical systems is much more difficult to achieve.

It is such
considerations that prompted the development of the program on ``Magnetic
Phenomena in Novel Materials and Geometries" this Fall at the ITP.
We report on the week-long conference that preceded it, which 
played a role in setting
directions for the longer program.
 
\section{Overview of Magnetism: Old and New}

A special session, devoted to longstanding central problems in magnetism 
with continuing importance and excitement for current and future
science, featured talks by
Fisher, Birgeneau, Affleck and Sawatsky. Fisher emphasized
the role of magnetism in the understanding of critical phenomena,
especially multicriticality and scaling. He discussed the
origin of bicritical and tricritical points in magnetic systems.
The variety of magnetic systems arising from changes in dimensionality,
anisotropy, magnetic field, ferromagnetic versus antiferromagnetic
couplings, incommensurate versus commensurate phases,
have provided tremendous insight into the subject of
phase transitions and classical critical phenomena. The 
antiferromagnetic next nearest neighbor Ising (ANNNI) model \cite{annni}
provides a very rich phase diagram, with multiple phases arising
out of a special frustrated point at zero temperature. These classical
studies may well be relevant to studies of multiple striped phases
in doped antiferromagnets, discussed later by Sachdev.

Birgeneau discussed neutron scattering
studies of two dimensional quantum Heisenberg
antiferromagnets and their comparisons with theory \cite{elstner}.
As is well known, the correlation length of the
spin-half cuprate materials is remarkably well
described by the quantum non-linear sigma (QNL$\sigma$) model
expressions of Chakravarty, Halperin and Nelson \cite{chn} and of
Hasenfratz and Niedermeyer (CHNHN) \cite{hn}. However, these expressions do
not adequately represent the experimental behavior for
quasi-2D antiferromagnets with $S>1/2$. This discrepency
is evident on comparison of the CHNHN
expressions with high temperature series
expansions and Quantum Monte Carlo simulations in the regime
where the correlation length is only a few lattice constants. 
It is now evident
that the QNL$\sigma$ correspondence is strictly valid only at
very low temperatures. At those temperatures systems with
larger spin can have rather large correlation lengths.
Experimental systems always have anisotropies and interplane
couplings, which will cut off the two-dimensional antiferromagnetic Heisenberg
description at sufficiently large length scales. 
Thus the QNL$\sigma$ description
is experimentally appropriate primarily for spin 1/2 systems.
Hasenfratz has recently discussed a way to incorporate the effects
of cutoffs into the theory \cite{hasen}, 
which allows a description of spin-dependent
correlation lengths at shorter length scales.

Birgeneau also discussed recent neutron scattering studies of 
Sr$_2$Cu$_3$O$_4$Cl$_2$ \cite{kim}. This quasi-two-dimensional 
material presents a
fascinating example of the ``order-by-disorder" phenomenon, whereby
quantum fluctuations lift accidental classical degeneracies
and cause spurious Goldstone modes to become gapped, leaving only 
true symmetry-related
Goldstone modes gapless. Neutron scattering
experiments have been used to study the excitation spectra of
this material, and have validated Shender's theory of quantum 
order-by-disorder. Recent theoretical studies of
Aharony, Harris and collaborators \cite{yildirim} on the selection of low
temperature ordering patterns in various cuprate materials
due to quantum fluctuations,
anisotropies, and weak interplane couplings further emphasize
the importance and generality of these order-by-disorder effects.

Affleck emphasized the importance of logarithms in the low temperature
properties of spin-chains. With the development of field-theoretic
bosonization techniques, the one dimensional spin problems are
now very well understood, including various logarithmic corrections.
These logarithms arise due to variables 
which are marginally relevant in the renormalization group sense, 
and they appear 
in many different properties \cite{agsz}.  Perhaps the most important 
manifestation of such logarithms is the
temperature dependence of the uniform susceptibility \cite{eggert}, which 
approaches its zero temperature value with an infinite slope, rather than
continuing the apparently smooth higher temperature behavior first described by
Bonner and Fisher \cite{bonner1}. 

Logarithms have also played
a very important role in analyzing numerical data on spin chains.
In particular, they lead to slow convergence of numerical approaches.
One way to think of the problem is in terms of length dependent effective
exponents, which only slowly approach their asymptotic values.
However, once the effects of leading logarithms are taken into
account, the numerical results show remarkable consistency with
theory \cite{bonner2}. There have also been efforts to interpret experimental
data on uniform susceptibility and NMR relaxation rates in terms
of these logarithmic corrections. However, in these cases their 
usefulness is less convincing due to the possibly many other perturbations
such as impurities and interchain couplings, which can strongly influence
the low temperature behavior of these systems.

Sawatzky discussed the importance of orbital degeneracy in understanding
magnetic systems. This is an old field, where seminal work was done 
first by Jahn and Teller and later by
Kugel and Khomskii \cite{kugel}. Recent discoveries of new materials, novel
synchrotron based probes which can directly observe orbital ordering,
and many technologically 
important phenomena such as colossal magnetoresistance (CMR)
and spin-electronics where orbital degeneracies play a role, have
led to a resurgence of interest in these systems. Sawatsky emphasized
the importance of orbital degeneracy and ordering in such systems
as LiVO$_2$ and V$_2$O$_3$ \cite{hfpen}. Understanding many of these oxide
materials requires close interplay of electronic structure (LDA
and LDA+U versions of density functional theory) and many-body approaches 
(such as dynamical mean-field theory).

Sawatzky also pointed to some continuing puzzles in the planar
insulating cuprate materials. While the spin-$1/2$
nearest neighbor Heisenberg model,
and, in particular, its treatment via spin wave theory, has presented an excellent
quantitative picture for the low energy excitations in the cuprate
materials, there are still some mysteries related to high energy
excitations \cite{lorenzana}. First of all, the lineshape and polarization
dependence of Raman experiments, which measure
2-magnon spectra, are not fully understood. The same is true
for optical absorption experiments. Finally, even in neutron
scattering there is substantial spectral weight in multimagnon
excitations, as well as evidence that the one-magnon dispersion
deviates from the nearest-neighbor Heisenberg spectrum
at higher energies \cite{aeppli2}. 
All these deserve further theoretical attention.

\section{Quantum Phases and Quantum Critical Points}

One important issue discussed in the conference is related to
the nature of ``quantum phases" in insulating and doped
two-dimensional antiferromagnets \cite{sachdev}. As discussed by Sachdev,
these phases can be characterized by the symmetries which
are spontaneously broken. Three classes of broken symmetries were
considered and are known to arise in some parts of the
phase diagram of doped two-dimensional antiferromagnets.
 Antiferromagnetically
ordered phases break spin-rotational symmetry. The superconducting
phases break the $U(1)$ gauge symmetry, whereas dimerized
and striped phases break lattice translational symmetry. 
By generalizing the spin-rotational symmetry from $SU(2)$ to
$Sp(N)$, and by studying the large-$N$ limit, Sachdev and Vojta \cite{vojta}
have shown that multiple striped phases, with varying periodicity
are possible upon doping, before the system turns into a
d-wave superconductor. Striped phases have been observed in many
high temperature superconducting materials. They have been argued
by Emery, Kivelson and collaborators \cite{kivelson}
to result from a competition between
tendencies for phase separation and long-range coulomb repulsion.
On the other hand, striped phases have also been observed
recently in density matrix renormalization group studies
of t-J models by Scalapino and White \cite{white}, without long-range coulomb
interaction and in parameter regimes where there is no
phase separation. The mechanism for stripe formation and the interplay
of stripes and pairing 
remains one of the key issues in the field \cite{hellberg}.

The divergence of the correlation length of a system  only in the limit 
$T\to 0$  has been designated ``quantum critical" behavior. A ``quantum
 critical point" separates two distinct zero temperature phases
(e.g., one magnetically ordered and the other disordered, or one
metallic and the other insulating) as some parameter is varied.   At these
critical points, one expects the temperature to set the 
characteristic energy scale for the system and  dynamical susceptibilites
to show scaling in temperature and frequency \cite{chn,csy}. 
Starykh discussed one
of the best studied quantum critical systems, namely the spin 1/2
Heisenberg chain. In this case the scaling behavior is complicated
by logarithms \cite{starykh}, 
which he argued is essential for understanding recent
NMR experiments by Takigawa \cite{takigawa} on the material Sr$_2$CuO$_3$. 
Aeppli presented
an example of a quantum critical point in an
itinerant-magnetic heavy fermion material \cite{aeppli},
where at the critical point the system shows non-Fermi liquid
behavior. The dynamical susceptibility shows very simple scaling
in temperature and energy. However, the associated critical exponents
are not well understood. Imai discussed NMR measurements in the
cuprate materials \cite{imai} as one goes from the insulating antiferromagnetic
phase to the metallic or superconducting phases upon doping. He pointed
out missing intensity in his NMR signals at low temperatures,
which he interpreted as evidence for stripe formation. Clearly,
quantum phase transitions and quantum critical points
in itinerant magnetic systems require further study.

\section{Frustration}

Strongly frustrated magnetic systems provide another class of problems  
that are not yet well understood. Frustration can have many
origins. It can arise from competing ferromagnetic and antiferromagnetic
exchange interactions placed randomly in a system as in conventional
spin-glasses, or from antiferromagnetic
interactions between spins on odd-length loops as happens
in triangular and Kagom\'e lattice antiferromagnets, or
from competition between exchange anisotropy and field
terms as in the transverse Ising model. It can also come about from
competition between superexchange and double exchange terms
which favor different spin alignments,
or from multiple spin exchange processes of odd
and even number of spins. One interesting consequence of this
exchange is enhanced entropy at low temperatures and what was
termed by Ramirez a ``spectral weight downshift'' \cite{moessner}.  A beautiful
example of this is the ``spin-ice'' system Re$_2$Ti$_2$O$_7$ \cite{ramirez},
which has a low temperature entropy similar to that associated
with the positions of hydrogen in ice as first discussed by
Pauling. 

Another general consequence of frustration is that
fluctuations play a very important role in selecting the
ordered state --- the phenomenon of order-by-disorder
discussed above. Furthermore, in systems with strong quantum
fluctuations, accidental degeneracy (beyond what symmetry would
dictate) can be lifted by superposition
of many states in a resonating valence bond scenario first
introduced by Anderson \cite{anderson}. In this case, one expects a gap in
the excitation spectrum and possibly excitations with exotic
quantum numbers. The talks by Lhuillier and Mila emphasized
various quantum spin systems which show spin-gap behavior. In particular,
they discussed the spin-half Heisenberg model
on the kagom\'e-lattice, 
which has intriguing properties. There appears to be
a gap in the spin-excitation spectra, but there are many
low-lying singlet excitations. The number of singlet states
below the lowest triplet appears to grow exponentially with
the size of the system \cite{lhuillier}. 
Mila \cite{mila} presented a theoretical framework
for understanding these low lying excitations.
Chubukov discussed order-by-disorder phenomena in double-exchange magnets,
and presented his results in context of the CMR manganite materials.
Moessner discussed the Ising magnets in a transverse field
on triangular and kagom\'e lattices, and argued that the triangular-lattice
model is ordered whereas the kagom\'e lattice model is not.
These systems deserve further theoretical attention.

\section{Spin Gaps}

The problem of spin gaps has attracted considerable attention recently
due to the synthesis of a large number of new materials which exhibit
behavior characteristic of such gaps, notably thermally activated  
magnetic susceptibilities. As was pointed out by Khomskii, there are many
routes to spin-gap behavior, in systems with spin-rotational
symmetry. As proposed first by Haldane \cite{haldane},
it is now established that spin gaps are generic
to quasi-1D Heisenberg spin systems with integer spin per unit cell.
A spin gap is also natural when the system consists of
finite clusters of spins which have singlet ground states, which
are then weakly coupled to other clusters. Other examples  
include spin-Peierls, orbitally degenerate, and strongly frustrated
spin systems. 

Reich discussed several quasi-one- and two-dimensional 
experimental materials which are
strongly dimerized, whereas Kodama presented experimental
results on the quasi-2D spin-gap materials CaV$_4$O$_9$  \cite{cav4o9} and 
SrCu$_2$(BO$_3$)$_2$ \cite{srcubo}. Neutron scattering is the most natural
tool for studying the spin dynamics of these systems due to
its detailed frequency and wavevector resolution. Reich showed
that sum rules and single-mode approximations often provide
quite accurate quantitative description of the spin dynamics
in strongly gapped systems \cite{reich}.
The material SrCu$_2$(BO$_3$)$_2$ discussed by Kodama
is particularly interesting from a theoretical point of view
because the spins have an exactly known quantum mechanical ground state 
\cite{shastry}, its
exchange constants put it
close to a quantum critical point and it exhibits magnetization plateaus
as a function of applied magnetic field \cite{ueda}.

Kotov discussed the nature of elementary excitations and bound
states in spin-gap systems \cite{kotov}.
Khomskii and Poilblanc focussed on spin-Peierls systems, especially
the inorganic spin-Peierls material CuGeO$_3$ \cite{poilblanc,mostovoy}. 
Both interchain couplings and dynamical phonons are important
in understanding the properties of these materials.
As was evident from these talks, the existence of soliton-like
excitations and their bound states is an exciting topic of current
research.

\section{Spin Chains and Spin Ladders}

Quantum spin chains have long been a pet subject of theorists, as
they are mathematically more tractable than higher dimensional systems
and exhibit various exotic
many-body phenomena. Yet many important results regarding even
the spin 1/2 chain have only recently been obtained. Affleck,
Eggert and Takahashi \cite{eggert} found that the uniform susceptibility
approaches its T=0 values with an infinite slope, as mentioned
above. With the
help of the Bethe ansatz, the uniform susceptibility of the
spin 1/2 chain is now known very accurately at all temperatures.
Johnston described ongoing efforts to develop accurate fits to the 
susceptibilities and specific heats
of uniform and alternating spin 1/2 chains and ladders, to make
detailed comparisons between theoretical models and experimental
results \cite{johnston}.

Spin ladders allow one to interpolate, by increasing the number of
legs, between one and two dimensions \cite{dagotto}.
The existence of Cuprate materials which exhibit ladder-like magnetic
structure and behavior have 
further increased interest in these systems. Sierra described
variational approaches to studying ladders \cite{sierra}, whereas Solyom
discussed different massive and critical phases in these systems \cite{solyom}.
Cabra explained the appearance of magnetization plateaus \cite{cabra} in 
quasi-1D spin systems as arising from the formation of
strongly correlated states at finite magnetizations. Weakly
coupled arrays of spin 1/2 chains were discussed by Sandvik \cite{sandvik}.
Using a quantum Monte Carlo simulation and a chain mean-field
theory, he was able to show that such systems are long range
ordered. The question of frustration in such weakly coupled
spin chains remains to be explored.

\section{New Magnetic Materials and Phenomena}

Advances in magnetism have long been driven by new materials,
from the ancient discovery of native lodestone to more recently
studied systems exhibiting novel phenomena such as spin-Peierls behavior. 
The discovery of interesting new materials continues
at a high rate, certainly within the fertile class of oxides 
which have supplied more than their share of novelties, but also 
in a variety of other materials.

\subsection{Oxides of Transition Metals}

A new system discussed by Keimer is the pseudo-quintenary 
system\cite{keimer} 
(La$_y$Y$_{1-y}$)$_{1-x}$Ca$_x$TiO$_3$.  With varying $y$, which is
thought to tune the bandwidth (or on-site Coulomb repulsion to bandwidth
ratio U/W) there is an antiferromagnetic to ferromagnetic transition.
With varying doping level $x$ there is an insulator-to-metal transition
which also depends on $y$, as in the similar manganite system 
discussed in Sec. IX.  This formally Ti$^{3+}$, $d^1$ 
configuration appears to be a prime one for orbital ordering\cite{kugel},
yet structural studies indicate
that in the La(Y)-rich regime the O$_6$ octahedron 
is undistorted, and there
is no change at the magnetic ordering temperature.  YTiO$_3$, on the
other hand, is Jahn-Teller distorted and orbitally ordered, as reproduced
in the calculations of Sawada and Terakura.\cite{sawada}
The ions at the ends of the transition metal row of the periodic table
often show extreme behavior (recall, e.g. 
that only doped Cu$^{2+}$ systems have to date been reliably shown to
become high temperature superconductors),
so the observed special features of this system may reflect new physical 
phenomenon.

\subsection{Molecular Magnet Crystals}
Landee presented an overview of molecular-based magnets, which are either
free of oxygen or for which oxygen has no active role.  While organic
ferromagnets exist, these molecular magnets typically consist of 
inorganic magnetic molecules which may be sheathed in organic material.
An important feature of these materials is their unusually small 
exchange coupling, weak enough for reasonable 
applied magnetic fields to have dominating effects.\cite{landee}
The Mn$_{12}$O$_{12}$ system (``Mn$_{12}$'' (crystalline
Mn$_{12}$O$_{12}$(CH$_3$COO)$_{16}$(H$_2$O)$_4$)
has become a miniclassic, displaying 
jumps in magnetism in an applied field\cite{Mn12jumps} and unusual 
relaxation behavior\cite{Mn12relax} that
have become an active area of theoretical study.  The CuPzN (copper
pyrazine nitrate) system forms a clean, 1D Heisenberg S=1/2
system; however, the ordering temperature T$_N$=70 mK makes it difficult
to study at ``low temperature'' (T$<<$T$_N$).

The observation of quantization of the magnetization as a function
of magnetic field, believed to result from tunneling between different
quantum states, in systems in
which the individual moments are quite large and therefore expected
to behave classically,
forms the basis of a field now called quantum tunneling in molecular
magnets.  As described by
Friedman, spin has no explicit kinetic degree of freedom, and the
connection between the classical and quantum description must be
addressed.  Fortunately, very simple and clean samples can be produced
that are comprised of identical nanoscale molecular magnets such as
Mn$_{12}$ or ``Fe$_8$'' (a
similarly complex material with periodically placed magnetic molecules).

Resonant tunneling of the magnetization in molecular magnets results
from tuning of energy levels of different S$_z$ states. 
This phenomenon involves macroscopic relaxation 
by accumulation of microscopic processes.\cite{restunnel}  
In many contexts the large local moments (typically S$\sim$10) 
would put one well within
the classical regime, but the r\^ole of 
quantum effects is believed to be important and is under active study.
Understanding of the thermodynamics
and dynamics of these molecular magnets depends on the identification
of the symmetry breaking (or lowering) terms in the spin Hamiltonian
that governs their behavior.

\subsection{Heavy Fermion Metals}

The ``heavy fermion'' phenomenon has been known for two decades. Metals 
such as UPt$_3$, UBe$_{13}$, and CeRu$_2$Si$_2$, show very unusual 
temperature dependences\cite{stewart}
of resistivity, susceptibility, and heat capacity but finally settle into
a vastly enhanced Fermi liquid regime where the carriers have masses of
up to 1000 times the free electron mass.  It was tempting to 
interpret the behavior as simply that
--- an enhanced Fermi liquid, and nothing else. But why does a one
particle s-f hybridization model seem to work so well?  Why don't the
strong correlations induce Mott localization?  Aeppli presented data
which indicates that, at least in some cases, much more is going on.
In some of these systems, such as CeCu$_{6-x}$Au$_x$, the system can be
tuned to drive the characteristic ``degeneracy" temperature  
to zero.\cite{aeppli}  This brings in quantum critical behavior, so
that the susceptibility becomes simple to model, scaling so
as to collapse onto a single universal curve, but one described by
a non-analytic function whose origin is still a mystery. Moreover, 
anomalies are seen at all wave vectors, not just that of the antiferromagnetic
order parameter.  Kondo singlet unbinding appears to be occuring 
simultaneously with antiferromagnetism at the quantum critical point.

\section{Half Metallic Magnets}
The phenomenon of half metallic ferromagnetism was a theoretical discovery
by band theorists in the early '80s.\cite{degroot} 
In such a system, in which up spin
and down spin spectral densities are inequivalent, one spin direction 
has gapless charge excitations while 
the other spin direction has
a gap in its charge excitation spectrum.  Hence one channel is 
non-conducting while the other is metallic, which has been
dubbed ``half metallic.''  Many 
such ferromagnets (and ferrimagnets) have been predicted, particularly
in the Heusler and half-Heusler compounds, but also in perovskites
and spinels (Fe$_3$O$_4$).  One of the simplest, structurally and 
electronically, is CrO$_2$.  In spite of quite a bit of work, however, it
has been difficult to obtain conclusive evidence of half-metallicity in
these candidates.  The problem is traceable to the difficulty
in establishing 100\%
spin polarization of the charge carriers,\cite{soulen} 
either in the bulk of the material,
or after ejecting the carriers through a surface or interface, due to
extraneous spin scattering.   

This situation is beginning to change.  Park described spin-resolved
photoemission spectra (SRPES) on the CMR system La$_{0.65}$Sr$_{0.35}$MnO$_3$, 
predicted\cite{nrl} to be half metallic or very nearly so.  
With carefully prepared
(but not atomically flat) surfaces of a thin film, his group was able to
demonstrate\cite{park} that the film ordered magnetically 
(as seen in the SRPES spectra)
 at the
bulk Curie temperature, and that well below the Curie temperature only
electrons of a single spin direction were emitted at or near the Fermi
level.  Not only is this the first very strong evidence for a half metallic
system, but it also buttresses the widespread feeling that the phenomenon
of colossal magnetoresistance in the manganites is closely related to
its half metallic character.  It might be expected that bulk phenomena,
such as NMR relaxation times or the low temperature resistivity (due to
two-magnon processes), should provide telltale signs of half metallic 
character.  Furukawa indicated why this is not the case, as he argued
that $\rho \sim$ T$^{9/2}$ that has been quoted from Ogata's work in
the '60s becomes $\rho \sim$ T$^{3}$ for a more realistic model.\cite{furukawa}
This result can be taken as evidence for half metallicity in CrO$_2$.

\section{Spin/Orbital/Lattice Coupling}

The observation of ``colossal magnetoresistance'' (CMR)\cite{CMR} 
in the manganites,
where there is a insulator-to-metal transition at the
magnetic ordering (Curie) temperature, has been found to involve
several processes, which may be competing or symbiotic.\cite{review}  
A fundamental 
link between transport and magnetic order is provided by Zener's
``double exchange'' mechanism,\cite{zener} whereby an electron can hop between
neighboring magnetic (Mn) ions only if their spins are parallel.
The gain in kinetic energy due to the hopping becomes a driving force
for ferromagnetic ordering.
Experimentally, it has become abundantly clear that the manganites
are much more complicated than this; for example,
magnetic ordering (and CMR)
can be driven by ionic size variations at constant doping level, 
and structural transformation can be driven even by magnetic field alone.
These effects are two manifestations of strong magnetostructural
coupling.

Since the carriers in the doped manganites ({\it e.g.} 
La$_{1-x}$Ca$_x$MnO$_3$) occupy the doubly degenerate {\it $e_g$} orbitals
of Mn, the question of Jahn-Teller distortion arises, and the 
Kugel-Khomskii mechanism of ``orbital ordering'' comes into play.
The Mn-O bondlength distortions of LaMnO$_3$ and the resulting magnetic
structure (anti-alignment of ferromagnetically aligned layers)
can be understood in terms of this mechanism.   
Maekawa presented a theory of coupled spin and orbital degrees of
freedom within a double exchange model, with results that connect 
orbital ordering tendencies to large response
(such as CMR).\cite{maekawa}  He illustrated the 
competing tendencies with phase
diagrams where orbital ordering occurs at $x$=1/8, ferromagnetism peaking at
$x$=1/2, and antiferromagnetism returning at larger $x$, 
results that show similarities to
observed behavior.

Cheong focused on the observed phenomena at $x$=3/8, which is where
the Curie temperature peaks in the CMR systems.  Recent work indicates
that the system at this concentration phase separates, one phase being
metallic and ferromagnetic and the other non-conducting and microscopically
charge ordered (stripes separated by a few lattice constants).\cite{cheong}
Due to the strength of the long-range Coulomb interaction charge neutrality
can't be broken on the larger scale of the size of
the domains (0.1 to 1 $\mu$m),  
so both phases must exhibit the global value 3/8 of hole
concentration.  The insulator-to-metal transition appears to be a
percolative phase change.  Conductivity is through the ferromagnetic
metal domains, whose size can be varied by {\it chemical} pressure with
the substitution of Pr for La, but with constant carrier concentration
$x=3/8$.

The $x$=1/3 regime, where the CMR effect is strongest, was discussed by
D. Singh.  At low temperature this is a very good ferromagnetic 
metal,\cite{djs1}
calculated to be nearly half-metallic.  Its structure, both observed
and computed, has no Jahn-Teller distortion,
consistent with the fact that orbital order is destroyed by 
easy carrier hopping.  As the temperature is raised near and beyond the
Curie temperature, where hopping becomes increasingly more difficult
due to spin disorder, the observed crystal
structure shows no noticeable 
change.  The strong electron-lattice coupling that is so evident at
lower doping levels seems to be strongly suppressed\cite{djs2} 
in the CMR region
of the phase diagram.

\section{Spin Control}

\subsection{Spin Electronics}
Manipulation of the spin degree of freedom of conduction electrons 
leads to a new form of electronics, now dubbed
{\it spintronics}. This form of current and voltage control uses 
low resistance (hence low voltage and low power consumption) 
magnetic metals rather
than high resistance (high voltage) semiconductors such as Si. 
The spin-polarized current also offers entirely new
possibilities, such as manipulations of electronic signals by 
magnetic fields or vice versa, or novel effects in
ferromagnet/superconductor/ferromagnet sandwiches or multilayers. 
Quantum information storage and quantum
computation\cite{kane} are related phenomena that require further study. 

A description of the mechanisms and structures of some novel 
magnetoelectronic devices was provided by M. Johnson.  He described 
two devices: (1) the
magnetoquenched superconducting valve, in which a superconducting
electronic element is bathed in a fringing magnetic field\cite{mj1} whose
direction is readily controlled by an external field, 
and (2) the ``hybrid Hall device'' 
(hybrid ferromagnet-semiconductor nonvolatile gate), which appears
promising as a high density, low power nonvolatile memory device.\cite{mj2}   
Johnson also discussed the detection of the degree of spin polarization
by detecting the chemical potentials of the ferromagnet at an
interface with a semiconductor.

\subsection{Spin Coherence in Semiconductors}

The issue of spin polarized carriers in semiconductors has become
active due to interest in their possible use in electronics,
computing, and information storage.\cite{kane}  Awschalom described how 
laser pulses can be used to excite, and then probe, polarized
carriers in GaAs.  The precession of the spins in an
applied field can be used to monitor the spin coherence, which can
be unexpectedly long in time and far in space.\cite{awschalom}  
Lateral transport of
spins of up to 100 nm without loss of polarization has been observed.
Dramatic effects of magnetic field on spin relaxation have also been
seen.  In zero field, up and down spins relax equally; upon application
of a field, the lifting of the Zeeman degeneracy gives rise to a
varying polarization that reflects quantum beating between the
Zeeman-split spin levels.  A detailed theory of these effects remain
to be developed.

\section{Closing Remarks}

In 1982 Hurd commented\cite{hurd} that ``magnetism in solids {\it used to be}
a tidy subject'' (emphasis added).  It seems now that 
it can be said that, compared to now,
magnetism in 1982 was a relatively tidy subject.  Problems that were in
some sense solved (heavy fermions) have not stayed solved, while new
questions continually arise.  Some longstanding
problems may have been put to rest --
the susceptibility of the spin-half 1D Heisenberg model, for example.
Meanwhile, for each question that is settled, a number of new ones seem
to appear.  
\vskip 3mm
{\it Copies of the speakers transparencies, audio recording of
the presentations and a list of review articles and books on the
subject can be found at the ITP web address} 
http://www.itp.ucsb.edu {\it (follow links on magnetism).}
\vskip 2mm
This work is supported in part by the National Science Foundation
Grant PHY94-07194.

\end{document}